\documentclass[journal,transmag]{IEEEtran}
\usepackage{amsmath}
\usepackage{amsthm, amssymb}
\usepackage{color}
\usepackage{pdfpages}
\usepackage{subcaption}
\usepackage{cite}
\usepackage{fancyhdr} 
\usepackage{graphicx}
\usepackage{float}
\usepackage[sc]{mathpazo}
\usepackage{bm}
\usepackage{varwidth}
\usepackage{xcolor}
\usepackage{tikz}
\usepackage[hidelinks]{hyperref}

\definecolor{lime}{HTML}{A6CE39}
\DeclareRobustCommand{\orcidicon}{%
    \begin{tikzpicture}
    \draw[lime, fill=lime] (0,0) 
    circle [radius=0.16] 
    node[white] {{\fontfamily{qag}\selectfont \tiny ID}};
    \draw[white, fill=white] (-0.0625,0.095) 
    circle [radius=0.007];
    \end{tikzpicture}
    \hspace{-2mm}
}

\foreach \x in {A, ..., Z}{%
    \expandafter\xdef\csname orcid\x\endcsname{\noexpand\href{https://orcid.org/\csname orcidauthor\x\endcsname}{\noexpand\orcidicon}}
}
\newcommand{\orcid}[1]{\href{https://orcid.org/#1}{\textcolor[HTML]{A6CE39}{\orcidicon}}}

\newcommand*\circled[1]{\tikz[baseline=(char.base)]{
            \node[shape=circle,draw,inner sep=1pt] (char) {#1};}}
\newcommand{\uproman}[1]{\uppercase\expandafter{\romannumeral#1}}

\begin{document}
\title{A Model for Multiple Metal Spheres in Oscillating Magnetic Fields using Displaced Dipoles}
\author{\IEEEauthorblockN{Florian B\"onsel\orcid{0000-0002-7193-9643}\IEEEauthorrefmark{1}, Alfred M\"uller\orcid{0000-0003-3037-4831}\IEEEauthorrefmark{2}, Rafael Psiuk\orcid{0000-0002-8744-2084}\IEEEauthorrefmark{1}}
\IEEEauthorblockA{\IEEEauthorrefmark{1}\small{Department of Self-Powered Radio Systems, Fraunhofer Institute for Integrated Circuits, 90411 N\"urnberg, Germany} }
\IEEEauthorblockA{\IEEEauthorrefmark{2}\small{PickWerk GmbH, 90491 N\"urnberg, Germany} }
\thanks{Corresponding author: Florian B\"onsel (email: boensefn@iis.fraunhofer.de).}}

\IEEEtitleabstractindextext{%
\begin{abstract}
In this article, we derive a magnetic dipole model for two identical, electrically conducting, and permeable spheres that are exposed to an oscillating homogeneous magnetic field. Our model predicts both amplitude and phase of the induced field outside the spheres. The description is provided for parallel and transverse excitation relative to the axis through the sphere centers. This geometric decomposition allows the application of arbitrary excitation field directions. Our approach is based on one dipole per sphere. The origins of these secondary dipole fields are proposed to be found at positions slightly displaced from the sphere centers to consider the mutual interaction. This displacement and the resulting phase of the dipole moments strongly depend on the distance between the spheres as well as on complex-valued first and second order response factors, which contain material properties and the oscillation frequency. 
We demonstrate the usefulness of our displaced dipole model in terms of efficiency and accuracy compared to other computationally simple approaches.
\end{abstract}

}

\maketitle

\IEEEdisplaynontitleabstractindextext

\IEEEpeerreviewmaketitle


\section{Introduction}
\IEEEPARstart{W}{hen} metal objects are exposed to ac magnetic excitation fields, secondary magnetic fields in the object's neighborhood are induced from either magnetization of the material or eddy currents.
The individual properties of these secondary response fields are determined by material parameters, excitation frequency, object size, and geometry. Reliable predictions of this induced ac-behavior are important for many applications nowadays. One example is the detection of unexploded ordnance: the secondary field of hidden, buried metal objects is detected and the signals, based on the aforementioned features, are classified \cite{shubitidze2005fast}. Two other scenarios, where knowledge of induced magnetic fields is used, are sorting procedures \cite{o2017fast}, \cite{dholu2017eddy} or nondestructive testing \cite{hamia2014eddy}. Whereas in the first application the signals need to be distinguished against each other from a discrete set of objects, the latter makes use of anomalies in the spatial conductivity and permeability distribution of a material and thus measures the deviation from an expected response \cite{tsukada2018small, d2018fast}. Also in fundamental research studies, the reaction of objects to  magnetic fields is focused, like the self-assembly behavior \cite{messina2017assembly} or the particle motion in tissue \cite{valberg1987magnetic} induced by magnetostatic fields.

Most of such applications share two properties. Firstly, the recorded signals are used to extract information about the underlying measured object. Secondly, it is usually one object at a time which is excited by an external field, thus making it possible to neglect the interaction with other objects. 

However, in some applications, the magnetic impact of surrounding items needs to be taken into account either for minimizing parasitic effects of the setup environment \cite{shubitidze2003analysis} or to identify signals emerging from multiple coupled objects \cite{shubitidze2004coupling} with distances between them often larger than the object's dimensions.

Another conceivable application which uses the secondary magnetic field is the characterization and classification of ensembles of small objects. In these clusters, the objects have very small distances from each other. They can even touch their nearest neighbors. Due to additional degrees of freedom from the statistical distribution within the packing \cite{moghaddam2018rigid} and its size, new forward models can help to further study and pioneer signal classification from measured responses. 

Albeit numerical investigations received some attention, e.g. a powder of spherical particles in static magnetic fields \cite{bjork2013demagnetization}, analytic solutions of eddy current- and magnetization-based fields only exist for geometrically simple scenarios. Examples are an infinite metal plane or a single sphere in a homogeneous magnetic field \cite{kaden1959eddy}. 

To approach a forward model that predicts the magnetic response to external fields for general structures of closely packed objects, the reduction of the problem to the case of two interacting spheres is a good point to start, because it is the least complex and geometrically most simple scenario that still considers  mutual interaction. Some approaches which model the the arising field strength for two metal spheres already exist. Nevertheless, these are limited to either static and ac electric fields \cite{qizheng2002displaced, xie2016iterative} or static magnetic fields \cite{du2014micro} and thus narrow the range of extractable object information. For time-varying magnetic fields there is not only a spatial field strength distribution but also a phase relative to the oscillating excitation field that encodes the secondary field. Computationally simple approaches model the objects as magnetic dipoles. Typical dipole models \cite{mehdizadeh2010interaction, du2013generating, gao2015electric} are based on symmetric and homogeneous
magnetization and current distributions \cite{griffiths1962introduction}, an assumption which is violated when the spheres are close to each other and multipolar effects arise \cite{wang2003frequency, keaveny2008modeling, du2014numerical}. 
One way to take these effects into account is to introduce a spatial shift of the dipole field origins, as for example in \cite{qizheng2002displaced, du2014micro, vega2019self}. 

Finally, a computationally simple model that predicts both amplitude and phase of static and oscillating secondary magnetic fields and at the same time takes care of small sphere distances is missing. With this work, we aim to close this gap . 

The paper is structured as follows: in Section\,\uproman{2}, the theoretical foundation is laid by reviewing the analytic expressions for a sphere in an oscillating homogeneous and in an oscillating dipole field. Based on these, our proposed displaced dipole model is motivated and derived for two geometric cases in Section\,\uproman{3}. After discussing general features of the model, we provide a comparison with FEM simulation data in Section\,\uproman{4}. The work concludes with an evaluation of the new model, some notes on challenges, and further research ideas.
\section{Review of a single sphere in external magnetic fields}
\begin{figure}[b]
\centering
\includegraphics[width=0.4\textwidth]{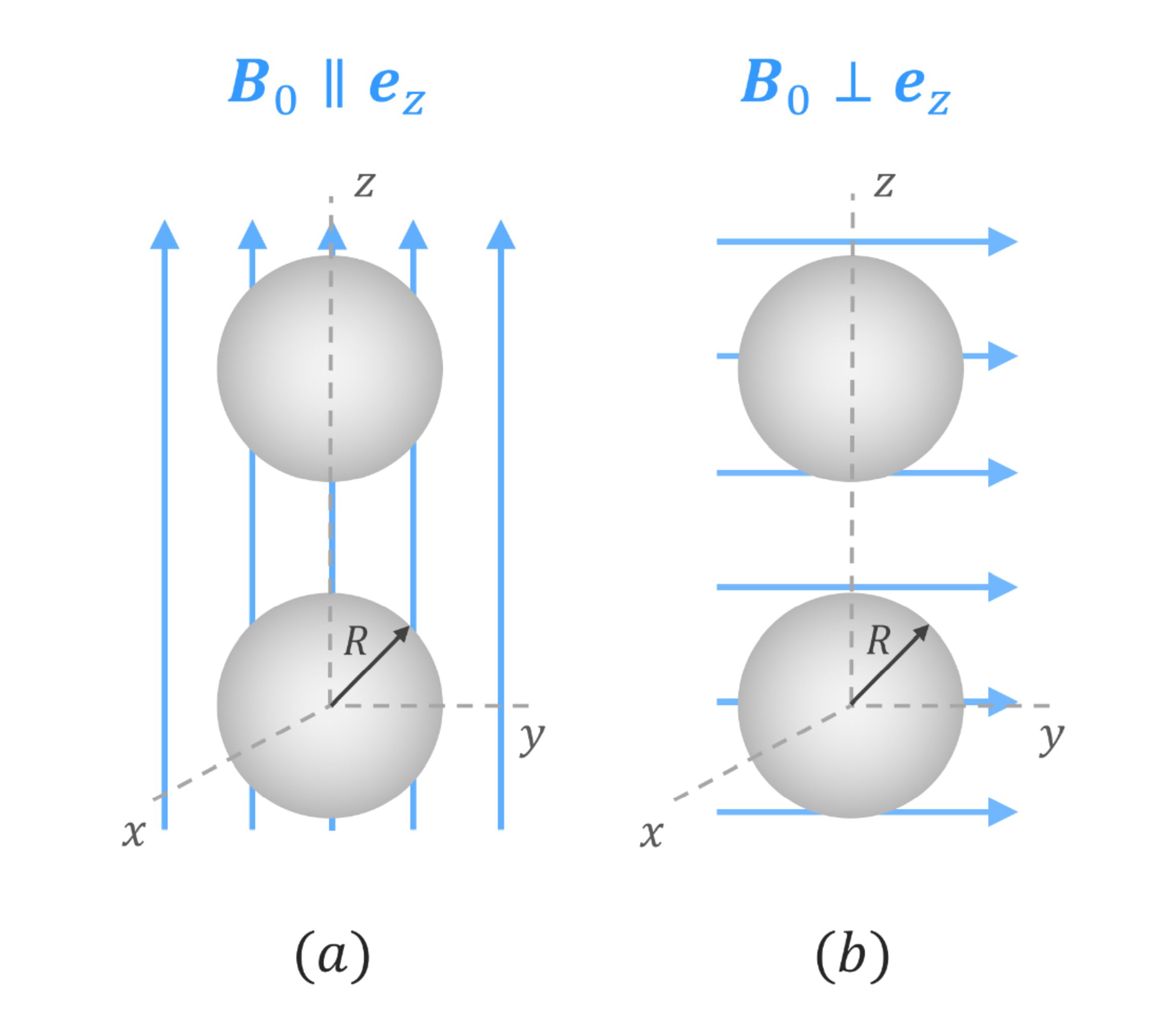}
\caption{\small{Geometric setting of the simultaneous excitation of two spheres. The figure illustrates parallel excitation (a) and transverse excitation (b).}}
\label{fig:parallel_transverse_geometry}
\end{figure}

We define the following geometric setup: two metal spheres with radius~$R$, electrical conductivity~$\sigma$, and relative permeability~$\mu_r$ are placed on the $z$-axis in a homogeneous magnetic field~$\bm{B}_0$  oscillating at frequency~$f$. As shown in Fig.\,\ref{fig:parallel_transverse_geometry}, the homogeneous excitation field can be written as $\bm{B}_0\!=\!B_0\bm{e}_i$ with $i\!=\!z$ or $i\!=\!y$, where the first case will be called parallel excitation, as in Fig.\,\ref{fig:parallel_transverse_geometry}(a), and the latter transverse excitation, as in Fig.\,\ref{fig:parallel_transverse_geometry}(b). 
An arbitrary direction of the homogeneous excitation field can then always be decomposed into these two cases. To mark the two geometric settings, the lower indices $\parallel$ and $\perp$ are used in this paper.

We introduce the necessary equations as a foundation for our model with a short review of the theory of a single sphere in external magnetic fields, following the work of \textit{Grant and West} \cite{grant1965interpretation}, pages 492 - 519.
\subsection{A metal sphere in a homogeneous magnetic field}
When a single sphere is excited by a homogeneous magnetic field $\bm{B}_0$, as it is shown in 
Fig.\,\ref{fig:parallel_transverse_geometry} for each of the two spheres, the secondary magnetic field outside such a sphere at position $\bm{r}$ coincides with a magnetic dipole field
\begin{align}
\bm{B}_D(\bm{r})=\frac{\mu_0}{4\pi r^3}\left[3\bm{e}_r(\bm{e}_r\cdot \bm{m}) -\bm{m} \right],
\label{eqn:dipole_field_general}
\end{align}
with $\bm{e}_r$ the unit direction vector from the sphere center to the field measurement point $\bm{r}$. The origin of this dipole field coincides with the sphere center.
The corresponding dipole moment is written as
\begin{align}
\bm{m}=-\frac{2\pi R^3 \alpha_1}{\mu_0}\bm{B}_0,
\label{eqn:secondary_dipole_B0}
\end{align}
where $\mu_0$ is the magnetic field constant and $\alpha_1\!=\!\left(X_1+jY_1 \right)\!\in\!\mathbb{C}$ a complex response factor, which depends on material parameters and frequency.
The time dependence of the oscillations $\mathrm{e}^{j2\pi t}$ can be ignored and only a relative phase, which is encoded in the complex value $\alpha_1$, needs to be considered.
\subsection{A metal sphere in the field of a magnetic dipole}
\begin{figure}[b]
\centering
\includegraphics[width=0.4\textwidth]{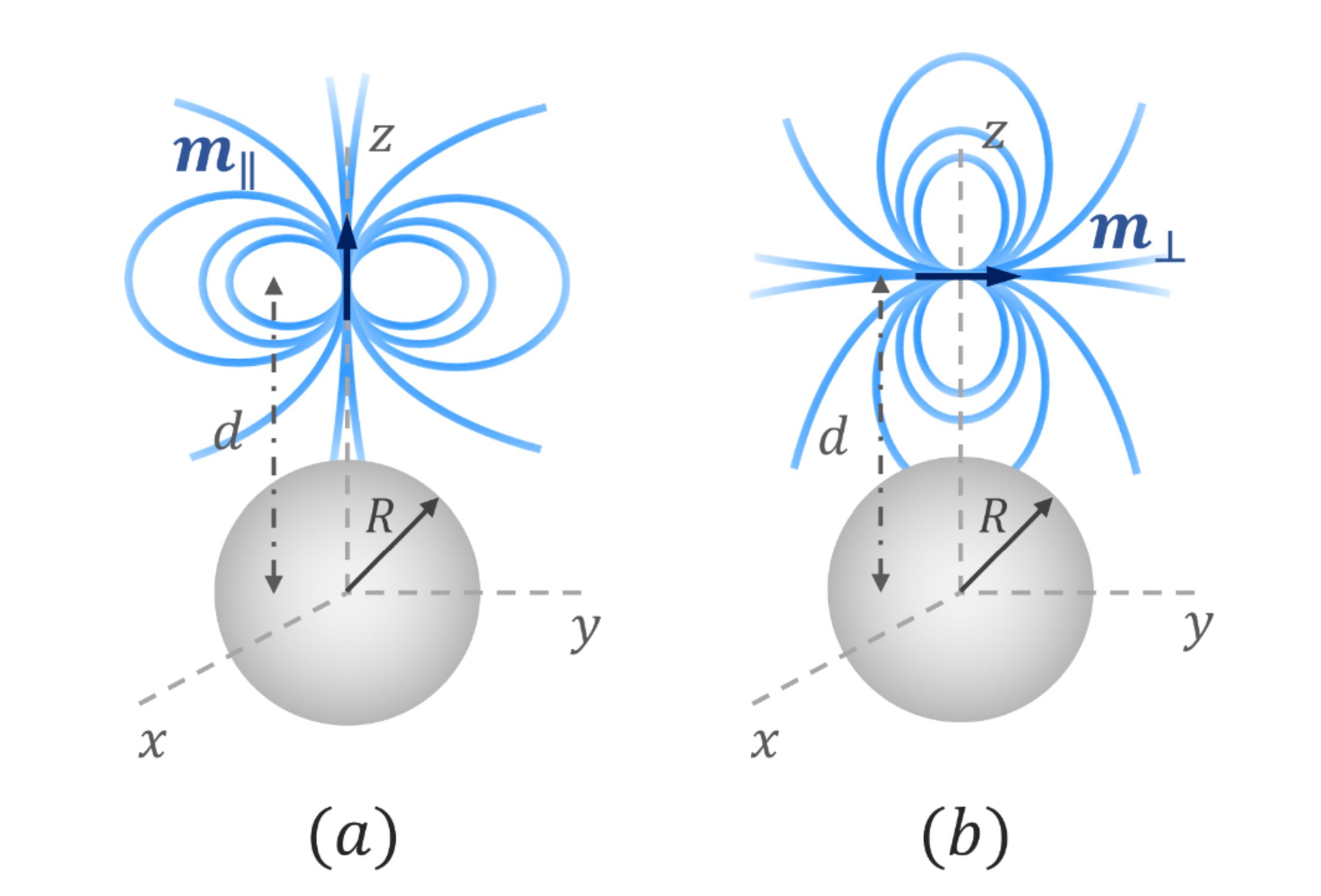}
\caption{\small{A metal sphere exposed to the field of an oscillating magnetic dipole on the $z$-axis, whose moment points either along the $z$-direction as $\bm{m}_{\parallel}$ (a) or along the $y$-direction as $\bm{m}_{\perp}$ (b). The oscillating dipole field is placed at distance $d$ from the sphere center.}}
\label{fig:sphere_in_dipole_field}
\end{figure}

In contrary to the homogeneous excitation, the same sphere can also be excited by the field of an oscillating magnetic dipole moment $\bm{m}\,=\,m\bm{e}_i$, which is placed w.l.o.g. at $(0,0,d)$, shown in Fig.\,\ref{fig:sphere_in_dipole_field}. Similarly to the above notation, $m\,=\,m_{\parallel}$ corresponds to $i\!=\!z$ and $m\,=\,m_{\perp}$ to $i\!=\!y$.
For the subsequent derivations and shorter notation, we only need the $y$ and $z$ components of the secondary field from this dipole excitation and evaluate these at points $(0,0,z)$. The field components are then written as
\begin{align}
B_{\parallel, z}(z)=-\frac{\mu_0 m_{\parallel}}{4\pi}\sum_{l=1}^{\infty}\frac{g_l}{z^{l+2}},
\label{eqn:H_r_parallel}
\end{align}  
for a parallel dipole excitation and as
\begin{align}
B_{\perp, y}(z)=-\frac{\mu_0 m_{\perp}}{4\pi}\sum_{l=1}^{\infty}\frac{h_l}{z^{l+2}},
\label{eqn:H_theta_perp}
\end{align}
for the transverse dipole excitation. The time dependence is again omitted. Both expressions (\ref{eqn:H_r_parallel}) and (\ref{eqn:H_theta_perp}) are power series, where the $l\!=\!1 $ terms are the dipole contributions and the higher terms are multipole contributions.
The factors $g_l$ and $h_l$ are defined as
\begin{align}
g_l = \frac{\alpha_lR^{2l+1}}{d^{l+2}} l(l+1),
\label{eqn:g_l}
\end{align}
and 
\begin{align}
h_l = \frac{\alpha_l R^{2l+1}}{d^{l+2}}\frac{l^2}{2}.
\end{align}
The expressions $\alpha_l=X_l+jY_l\,\in\,\mathbb{C}$ are complex response factors of order $l$. The first order factor also appears in (\ref{eqn:secondary_dipole_B0}). According to \cite{grant1965interpretation}, they are defined as
\begin{align}
\alpha_l=\frac{\left( \frac{1}{2} -(l+1)\mu_r \right)I_{l+\frac{1}{2}}(kR) +kR I'_{l+\frac{1}{2}}(kR)}{\left( 1/2+l\mu_r \right)I_{l+\frac{1}{2}}(kR) +kR I'_{l+\frac{1}{2}}(kR)},
\label{eqn:definition_response_factor}
\end{align}
with $k\!=\!\sqrt{j2\pi f \sigma \mu_r \mu_0}$. $I_{l+\frac{1}{2}}$ and $I'_{l+\frac{1}{2}}$ are the modified Bessel functions as well as their derivatives w.r.t. the argument $kR$.

\begin{figure*}[t]
\centering
\includegraphics[width=0.85\textwidth]{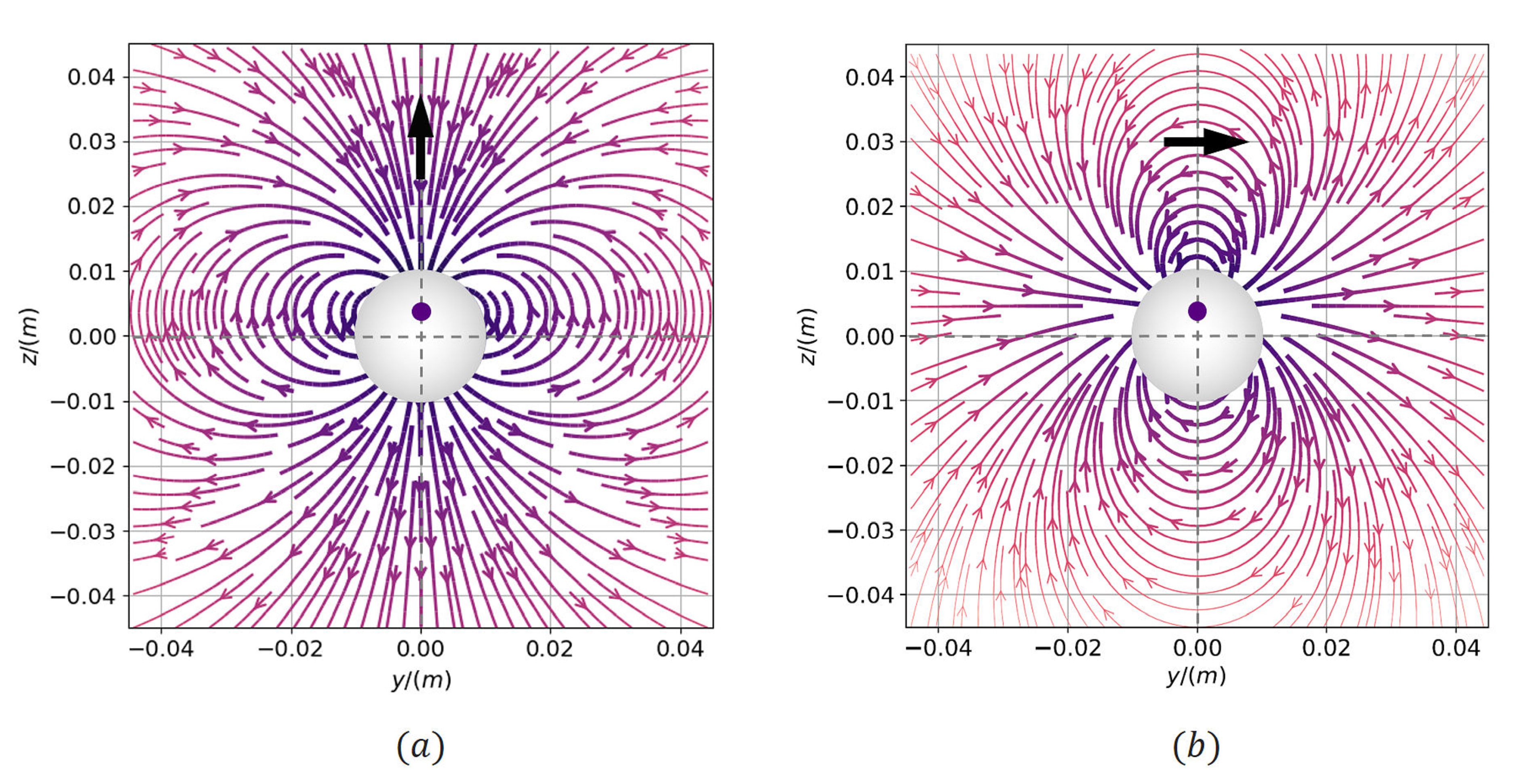}
\caption{\small{Field line distribution of the secondary magnetic field of a sphere, placed in the origin,  in the whole $y-z$ plane calculated with the equations from \cite{grant1965interpretation}. We set $m_{\parallel}=1$ for a parallel dipole excitation source (a) and $m_{\perp}=1$ for the transverse dipole excitation source (b) at $(0,0,3R)$ (big arrow). The plots show the case $f\,\sigma\!\gg\!1$. The origins of the resembling secondary dipole fields are indicated by dots and the lines have a logarithmic  scale in brightness and thickness according to the field strength.}}
\label{fig:field_distribution_GW}
\end{figure*}

We finish this section with a look on the geometric field distribution of the full equations, which are provided by \cite{grant1965interpretation} for any point in the $y-z$ plane, not restricted to points $(0,0,z)$ as in (\ref{eqn:H_r_parallel}) and (\ref{eqn:H_theta_perp}). This is shown in Fig.\,\ref{fig:field_distribution_GW}, where $f\,\sigma\!\gg\!1$.
To focus on the geometric distribution of the fields independent of the excitation source, we set $m_{\parallel} = m_{\perp}=1$. The field lines reflect the field strength in brightness and line width to a logarithmic scale. Dark and thick lines indicate large field strengths, whereas bright and thin lines mark  regions of weaker fields.

A main observation is that the multipole secondary field strongly resembles a dipole field. The origin of this secondary dipole field is indicated by a dot. It is slightly displaced from the center of the sphere along the axis which connects the sphere center and the origin of the oscillating excitation dipole field. As according to (\ref{eqn:H_r_parallel}) and (\ref{eqn:H_theta_perp}) the lower order multipole terms scale much stronger with the distance to the sphere center, we can only see fields which are dominated by the $l\!=\!1$ dipole terms.
The comparison between the factors of the power series gives another information which is $g_l\!=\!2\left(\!1\!+\!\frac{1}{l}\!\right)\! h_l$. This means that the secondary field around the sphere in  parallel dipole excitation is much stronger relative to the transverse case. For example, there is a factor $4$ between the dominating $l\!=\!1$ dipole terms and a factor $3$ between the next higher $l\!=\!2$ multipole terms.
From Fig.\,\ref{fig:field_distribution_GW}, this difference in the secondary field strength from dipole excitation between parallel and transverse case can be qualitatively seen by comparing the thickness of the shown field lines.
\section{The proposed displaced dipole model}
\subsection{The general procedure}
The following argumentation is provided only for the case of parallel excitation, shown in Fig.\, \ref{fig:parallel_transverse_geometry}(a). The steps for the transverse scenario are analog using (\ref{eqn:H_theta_perp}) instead of (\ref{eqn:H_r_parallel}). Therefore, the corresponding results for the latter scenario are provided in Section~\uproman{3},\,D without a fully repeated procedure.

\begin{figure*}[ht!]
\centering
\includegraphics[width=0.85\textwidth]{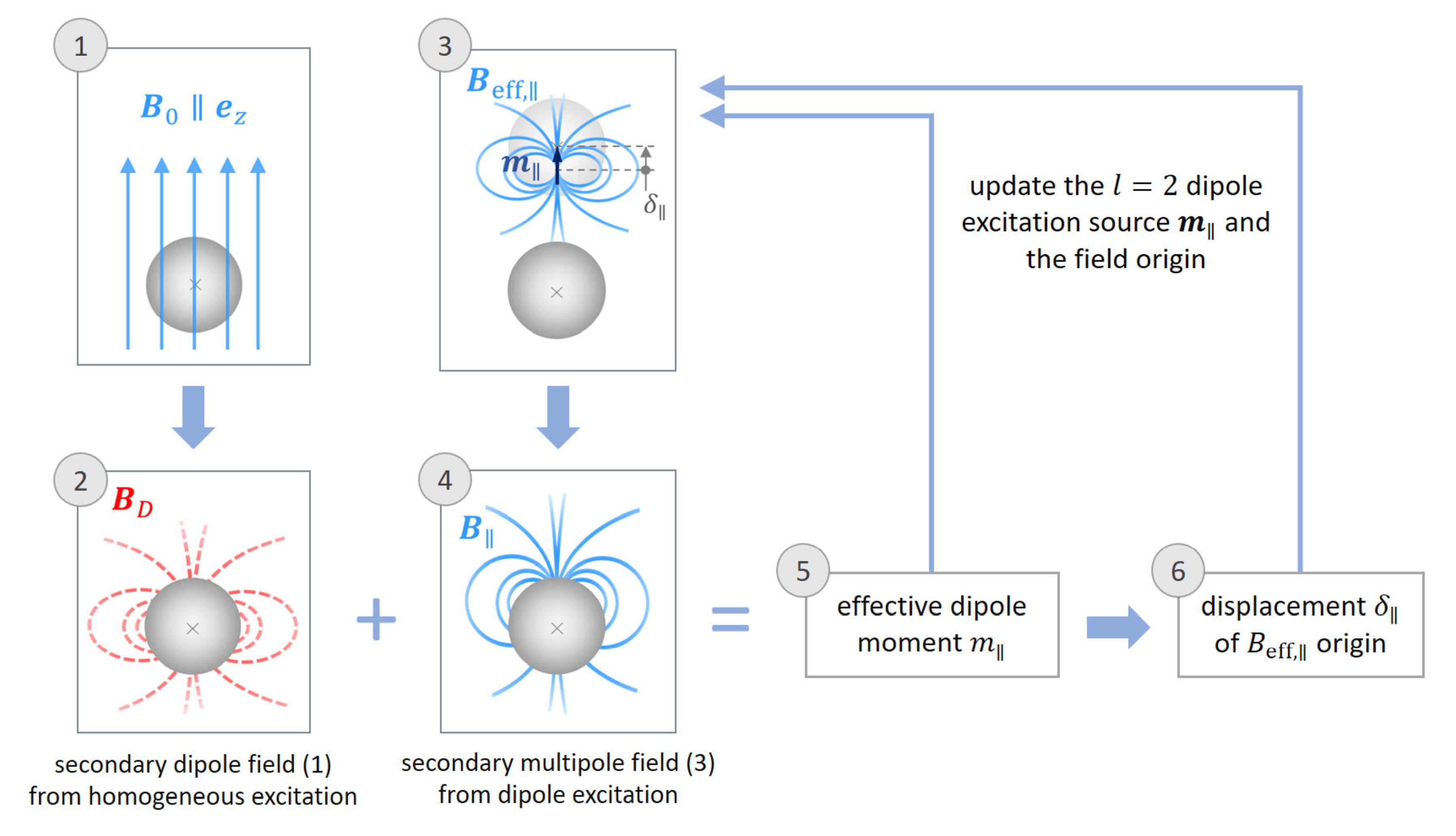}
\caption{\small{Block diagram of the procedure behind the displaced dipole model for the example of parallel excitation.}}
\label{fig:block_diagram}
\end{figure*}

The overall approach is illustrated in Fig.\,\ref{fig:block_diagram}, which shows a block diagram of the individual steps of our model for parallel excitation. Because of symmetry, it is sufficient to focus on sphere $1$, which we define to be in the coordinate center of Fig.\,\ref{fig:parallel_transverse_geometry},(a). The dipole moment of the upper sphere $2$ will be identical.

Firstly, each of the spheres in the defined setting from Fig.\,\ref{fig:parallel_transverse_geometry}(a), is excited by $\bm{B}_0$, shown in \circled{1}. As a consequence, a secondary magnetic field (\ref{eqn:dipole_field_general}) with its corresponding magnetic dipole moment (\ref{eqn:secondary_dipole_B0}) is induced for both of the spheres according to the theory of Section\,\uproman{2},\,A. This can be seen in block \circled{2}. Its origin is the sphere center.

From the fact that our approach assumes the final secondary field of the individual spheres as an effective dipole field, which we will call $\bm{B}_{\mathrm{eff}, \parallel}$, each of them is also exposed to $\bm{B}_{\mathrm{eff}, \parallel}$ of the other sphere, not necessarily with its origin in the sphere center. This is shown in\,\circled{3}. The induced field from this excitation can be modeled by (\ref{eqn:H_r_parallel}), which contains the full power series with all $l\!\geq\! 1$ terms. The resulting secondary field from this additional excitation corresponds to\,\circled{4}.
The sum of both secondary fields in \circled{2} and \circled{4} is the total secondary field of the single sphere, which acts back on the adverse sphere as an excitation again.
But to the best of our knowledge, there exists no analytic theory that describes the sphere secondary field due to field terms of order $l\!\geq\!2$, as they appear in (\ref{eqn:H_r_parallel}).
We therefore interpret the sum of \circled{2} and \circled{4} as an effective dipole field with dipole moment $m_{\parallel}$, corresponding to \circled{5}, and a field origin that is displaced from the sphere center by $\delta_{\parallel}$, corresponding to \circled{6}.
The result is then used to update the magnetic moment and the origin of the dipole excitation field in \circled{3}. From this point we repeat everything starting with \circled{1} and \circled{3} until we achieve convergence in the parameters $m_{\parallel}$ and $\delta_{\parallel}$.

Considering only the outcome in step \circled{2}, superimposing the two secondary dipole fields of both spheres, and stopping the procedure, is further referred to as the additive dipole model ($AD$). This solution ignores any further mutual interaction between the spheres, which would be considered in steps \circled{3} - \circled{6}, and will be used for comparison with the final model in Section\,\uproman{4}.

Summarizing, the bottom line of our proposed model is to stay on a formulation of dipoles, i.e. of order $l\!=\!2$ terms. From this, we expect deviations to the physical reality, as higher order terms are neglected. But this error is assumed to be balanced by the second part of our approach: the displacement of the dipoles. 
The complete model is titled the interacting displaced dipole ($IDD$) model.
\subsection{Effective dipole moments from mutual interaction}
This subsection yields the derivation of the effective dipole moment $m_{\parallel}$. We introduce the surface-to-surface distance $a$ along the $z$-axis. A distance $a\!=\!0$ means touching spheres. In\,(\ref{eqn:H_r_parallel}), the parameter $d$ describes the distance between the oscillating magnetic dipole moment as the excitation source and the center of the exposed sphere. Introducing a displacement $\delta_{\parallel}$ for the simultaneous excitation of two spheres, we place the dipole moments at  $\left(0,0,\delta_{\parallel}\right)$ for sphere $1$ and at $\left(0,0,2R+a-\delta_{\parallel}\right)$ for sphere $2$, such that $d=2R+a-\delta_{\parallel}$. Positive values of $\delta_{\parallel}$ describe  origins of the secondary magnetic fields that are closer to each other than the sphere centers.

We keep the displacement as a yet undefined variable and assume the following equation to hold at any point on the $z$-axis:
\begin{align}
\underbrace{\frac{\mu_0 m^{(n)}_{\parallel}}{2\pi \left(z-\delta_{\parallel}^{(n)}\right)^3}}_\text{\circled{5},\,\circled{6}} =
\underbrace{%
\vphantom{ \frac{\mu_0 m^{(n)}_{\parallel}}{2\pi (z-\delta_{\parallel}^{(n)})^3} } -\frac{R^3 B_0 \alpha_1}{z^3}}_\text{\circled{2}} + 
     \underbrace{%
\vphantom{\frac{\mu_0 m^{(n)}_{\parallel}}{2\pi (z-\delta_{\parallel}^{(n)})^3}}B_{\parallel,z}\left(m_{\parallel}^{(n-1)}, \delta_{\parallel}^{(n-1)}, z  \right)}_\text{\circled{4}},
\label{eqn:cond_parallel_dipole}
\end{align}
This equation, formulated with indices $^{(n)}$ by means of an iterative fixpoint solution, consists of the following parts: 

The left-hand side is the $z$-component of the effective secondary dipole field $\bm{B}_{\mathrm{eff}, \parallel}$, which is supposed to model the response of sphere $1$. It is equal to (\ref{eqn:dipole_field_general}) with magnetic dipole moment $\bm{m}\,=\,m_\parallel\,\bm{e}_z$ and a distance between the evaluation point $\bm{r}=(0,0,z)$ and the field origin of $z-~\delta^{(n)}_{\parallel}$, due to the proposed displacement, see \circled{5} and \circled{6} in Fig.\,\ref{fig:block_diagram}.
The first term on the right-hand side is the $z$-component of the secondary dipole field arising from the primary homogeneous excitation and corresponds to \circled{2} in Fig.\,\ref{fig:block_diagram}. Here, we insert (\ref{eqn:secondary_dipole_B0}) into (\ref{eqn:dipole_field_general}) and evaluate the field at $\bm{r}=(0,0,z)$, as here, the dipole field origin is the sphere center.
The second term is again the $z$-component of the secondary magnetic field (\ref{eqn:H_r_parallel}), which arises from the excitation due to the secondary magnetic dipole field $B_{\mathrm{eff}, \parallel}$ of the adverse sphere $2$. This contribution to the final secondary field of sphere $1$ corresponds to \circled{4} in Fig.\,\ref{fig:block_diagram}.

Equation (\ref{eqn:cond_parallel_dipole}) is an approximation and the dipole formulation that we use is only valid far away from the dipole source. Thus, we solve the equation in the limit $z\,\rightarrow\,\infty$ and get
\begin{align}
m^{(n)}_{\parallel}=-\frac{2\pi R^3\alpha_1}{\mu_0}\left[B_0 + \frac{\mu_0 m_{\parallel}^{(n-1)}}{2\pi \left(2R+a-\delta_{\parallel}^{(n-1)}\right)^3} \right].
\label{eqn:parallel_dipole_iteration_formula}
\end{align}
Of all $l\,\geq\,1$ terms, the iteration formula for the dipole moment only depends on the first order response factor $\alpha_1$. 
\subsection{Displacements as corrections to the dipole formulation}
\begin{figure}[t]
\centering
\includegraphics[width=0.35\textwidth]{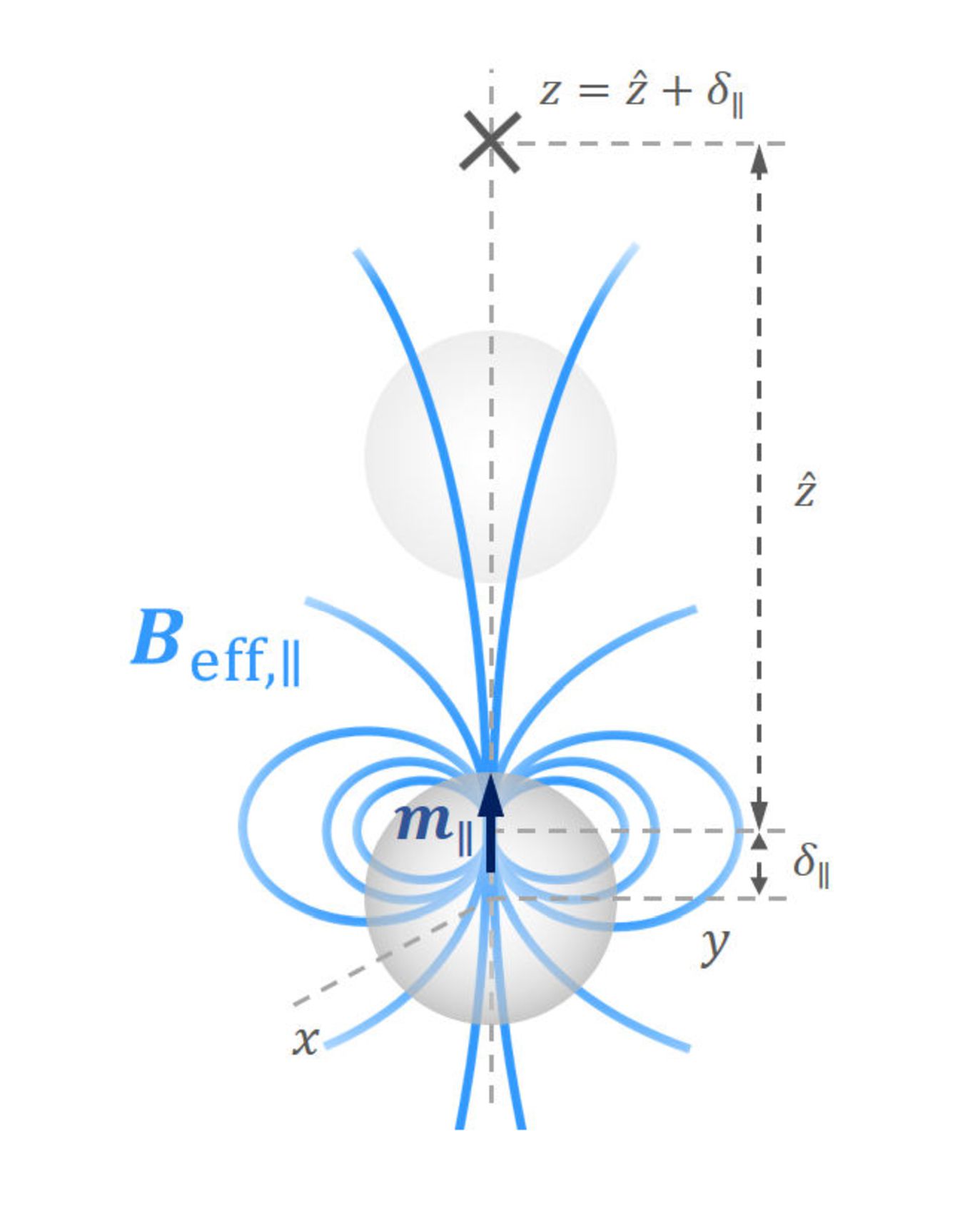}
\caption{\small{Localization of the secondary magnetic field origin. The field has the effective dipole moment $\bm{m}_{\parallel}$ and is assumed to originate at $\left(0,0,\delta^{(n)}_{\parallel}\right)$. The localization formula yields $\hat{z}$ as the distance from the measurement point at $(0,0,z)$ (marked with a cross) and the field origin.}}
\label{fig:Nara_picture}
\end{figure}

This part of the paper corresponds to \circled{6} in Fig.\,\ref{fig:block_diagram} and determines the above   introduced but yet unspecified displacement $\delta_{\parallel}^{(n)}$. As we argue that the field of each sphere can be approximated as a magnetic dipole field, we can use a technique proposed in~\cite{nara2006closed} for dipole field source localization. There, a compact formula for the distance between a field measurement point and the dipole source is presented by solely using data of the field strength and its spatial derivative at the field measurement point. The general formula provides the inversion of (\ref{eqn:dipole_field_general}) w.r.t. $\bm{r}$, which in our case is $\bm{r}=(0,0,z)$. 
For our geometric setting, the equation in\,\cite{nara2006closed} reduces to
\begin{align}
\hat{z}=-3\frac{B_{\mathrm{eff}, \parallel,z}(z)}{\partial_z B_{\mathrm{eff}, \parallel,z}(z)},
\label{eqn:Nara_equation}
\end{align}
where $\hat{z}$ is the distance between the magnetic dipole field origin and the field measurement point $z$, as illustrated in Fig.\,\ref{fig:Nara_picture}. The $z$-component $B_{\mathrm{eff}, \parallel, z}$ of the effective secondary dipole field in (\ref{eqn:Nara_equation}) is provided by the right-hand side of (\ref{eqn:cond_parallel_dipole}).
We find $\delta^{(n)}_{\parallel}=\!z\!-\!\hat{z}$. The localization formula is independent of the phase of the field. It is therefore sufficient to only use the absolute values $\left| B_{\mathrm{eff}, \parallel, z}(z) \right|$ and $\left|\partial_z B_{\mathrm{eff}, \parallel,z}(z) \right|$, but keep an additional minus sign in the denominator, that originates from the derivative of the $1/z^{l+2},~l\!\geq\!1$ terms.

For large distances $z$, we find
\begin{align}
\delta_{\parallel}^{(n)}=\lim_{z\rightarrow \infty} \left[ z - 3\frac{\left| B_{\mathrm{eff}, \parallel}\left(m_{\parallel}^{(n-1)}, \delta_{\parallel}^{(n-1)}, z  \right)\right|}{\left|\partial_z B_{\mathrm{eff}, \parallel}\left(m_{\parallel}^{(n-1)}, \delta_{\parallel}^{(n-1)}, z  \right)\right|} \right],
\label{eqn:limit_parallel_displacement}
\end{align}
which yields
{\small
\begin{align}
\delta^{(n)}_{\parallel}=-\frac{\Re\left\{ \rho^{(n-1)}\right\} \Re\left\{ \eta^{(n-1)}\right\} + \Im\left\{ \rho^{(n-1)}\right\} \Im\left\{ \eta^{(n-1)}\right\}}{\left| \rho^{(n-1)} \right| ^2},
\label{eqn:delta_parallel_final}
\end{align}}
with
\begin{align}
\rho^{(n-1)}=-12\frac{\pi R^3\alpha_1 B_0}{\mu_0} - 3m^{(n-1)}_{\parallel} g_1^{(n-1)},
\end{align}
and
\begin{align}
\eta^{(n-1)}= g_2^{(n-1)}m^{(n-1)}_{\parallel}.
\end{align}
The factors $g_{1,2}$ are now also equipped with an upper index, due to their dependence on the displacement of the previous step via $d=2R+a-\delta_{\parallel}^{(n-1)}$, see the definition (\ref{eqn:g_l}). 
We now have a set of iteration equations to find the dipole moments via (\ref{eqn:parallel_dipole_iteration_formula}), as well as the displacement via (\ref{eqn:delta_parallel_final}) for parallel excitation of two identical spheres. 

\subsection{Iteration formulas for transverse excitation}
For the transverse excitation of two spheres, as shown in Fig.\,\ref{fig:parallel_transverse_geometry}(b), the steps are analog. We only need to take the $y$-component of the fields on the $z$-axis and use
\begin{align}
\underbrace{%
\vphantom{ \frac{\mu_0 m^{(n)}_{\parallel}}{2\pi (z-\delta_{\parallel}^{(n)})^3} }-\frac{\mu_0 m^{(n)}_{\perp}}{4\pi \left(z-\delta_{\perp}^{(n)}\right)^3}}_\text{\circled{5},\,\circled{6}}=
\underbrace{%
\vphantom{ \frac{\mu_0 m^{(n)}_{\parallel}}{2\pi (z-\delta_{\parallel}^{(n)})^3} }
\frac{R^3 B_0 \alpha_1}{2z^3}}_\text{\circled{2}} + 
\underbrace{%
\vphantom{ \frac{\mu_0 m^{(n)}_{\parallel}}{2\pi (z-\delta_{\parallel}^{(n)})^3} }
B_{\perp, y}\left(m_{\perp}^{(n-1)}, \delta_{\perp}^{(n-1)}, z  \right)}_\text{\circled{4}},
\label{eqn:cond_perp_dipole}
\end{align}
instead of~(\ref{eqn:cond_parallel_dipole}). The numbers beneath the equation correspond again to diagram Fig.\,\ref{fig:block_diagram}, which can be drawn similarly for the transverse excitation and  the field component $B_{\mathrm{eff}, \parallel,z}$ needs to be replaced with the $y$-component $B_{\mathrm{eff}, \perp,y}$ for the localization formula (\ref{eqn:Nara_equation}). The iteration formulas for this case can then be written as
\begin{align}
m^{(n)}_{\perp}=-\frac{2\pi R^3\alpha_1}{\mu_0}\left[ B_0 - \frac{\mu_0 m_{\perp}^{(n-1)}}{4\pi \left(2R+a-\delta_{\perp}^{(n-1)}\right)^3} \right],
\label{eqn:perp_dipole_iteration_formula}
\end{align}
for the complex-valued dipole moment and
\begin{align}
\delta^{(n)}_{\perp} = \frac{\Re\left\{ \overline{\gamma}^{(n-1)}h_2^{(n-1)}m^{(n-1)}_{\perp} \right\}}{\left| \gamma^{(n-1)} \right| ^2},
\label{eqn:delta_perp_final}
\end{align}
with
\begin{align}
\gamma^{(n-1)}=-6\frac{\pi R^3B_0}{\mu_0} + 3m^{(n-1)}_{\perp} h_1^{(n-1)},
\end{align}
for the transverse displacement, where the bar on $\overline{\gamma}^{(n-1)}$ means the complex conjugate. The numerator in (\ref{eqn:delta_perp_final}) looks structurally different compared to the parallel displacement (\ref{eqn:delta_parallel_final}). This is a result from evaluating the corresponding limit (\ref{eqn:limit_parallel_displacement}) in the transverse scenario, where $\parallel$ is replaced by $\perp$.
\subsection{Dipole displacements in limit scenarios}
\begin{figure}[t]
\centering
\includegraphics[width=0.5\textwidth]{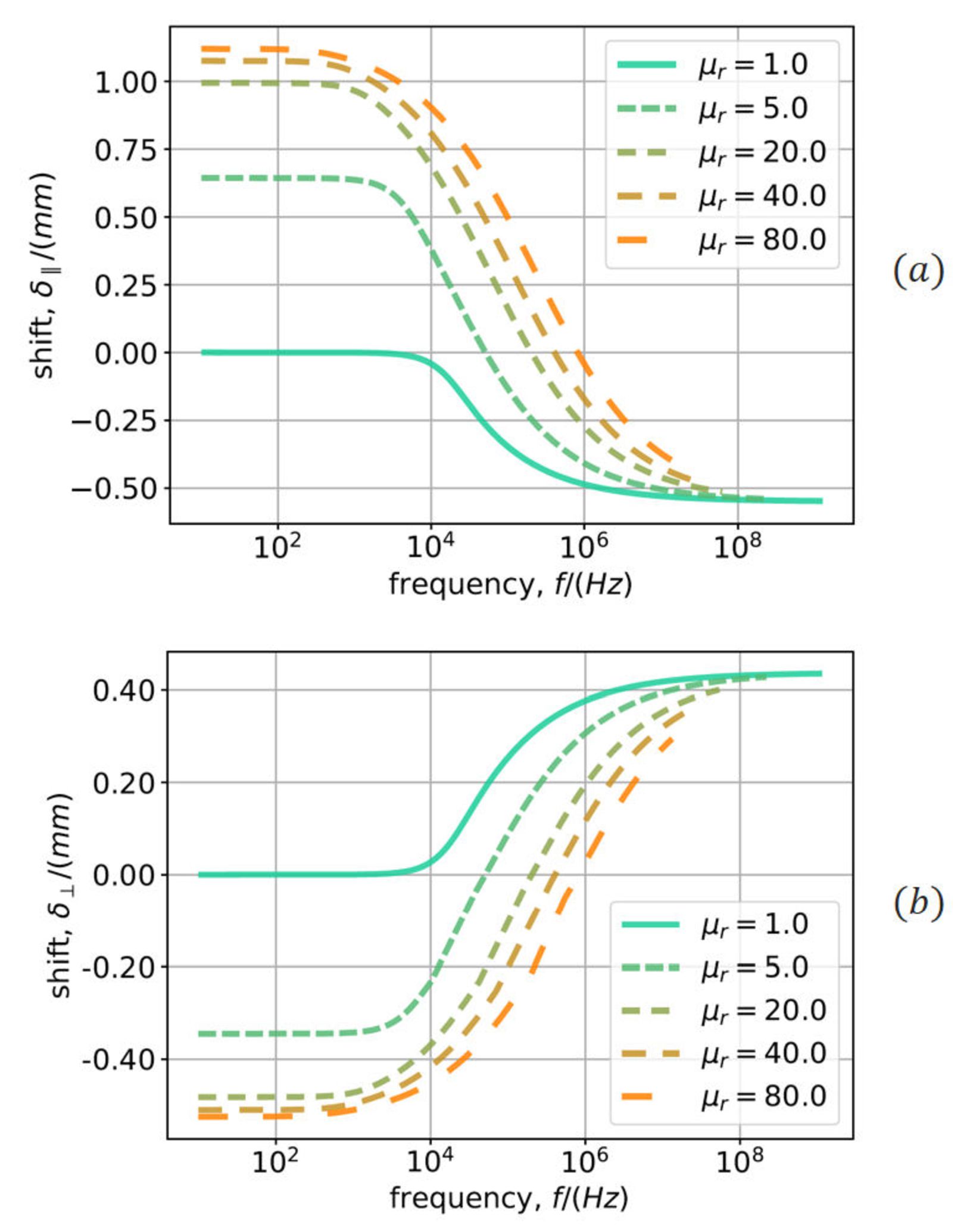}
\caption{\small{Parallel displacement $\lim_{n\rightarrow \infty}\delta_{\parallel}^{(n)}$ (a) and transverse displacement $\lim_{n\rightarrow \infty}\delta_{\perp}^{(n)}$ (b) against oscillation frequency of the excitation field for various values of relative permeability $\mu_r$. The parameters were set to $a\,=\,0.1\mathrm{mm}$, $R\,=\,10\,\mathrm{mm}$, and $\sigma\,=\,10^{6}\,\mathrm{S/m}$.}}
\label{fig:displacements}
\end{figure}

The frequency-dependent behavior of both $\delta_{\parallel}~=~\lim_{n\rightarrow \infty}\delta_{\parallel}^{(n)}$ and $\delta_{\perp}\!=\!\lim_{n\rightarrow \infty}\delta_{\perp}^{(n)}$ is shown in Fig.\,\ref{fig:displacements}, where we use two spheres of radius $R\!=\!10~\mathrm{mm}$, distance $a\!=\!0.1~\mathrm{mm}$, and $\sigma\!=\!10^{6}~\mathrm{S/m}$.
From Fig.\,\ref{fig:displacements}, the shifts for the low-frequency limit $f\,\rightarrow\,0$ can be identified as functions of $\mu_r$.

The signs of the displacements, i.e. $\delta_{\parallel}\!>\!0$ and $\delta_{\perp}\!<\!0$ for $f\!\rightarrow\!0$, are a result from the static excitation. The spheres are mostly magnetized, following the excitation field direction and the origins of the dipoles move to positions where the magnetization density is strongest. This is exactly the argumentation in \cite{du2014micro}, where two spheres are exposed to a static magnetic field. Another interesting result is that we find a convergence behavior of the displacements, i.e. a maximum absolute shift in the low-frequency case, even if the permeability is increased further.

The contrary extreme, indicated by values $f\sigma\!\gg\,1$, makes the displacements independent of the relative permeability. In this case, induced eddy currents are the main source of the secondary magnetic field instead of magnetization. Fig.\,\ref{fig:displacements} also shows that for $f\sigma\!\gg\!1$ the signs of the final displacements are reversed, i.e.~$\delta_{\parallel}\!<\!0$ and $\delta_{\perp}\!>\!0$. Whereas in \cite{du2014micro}, the static case is explained, a similar explanation can be made for higher frequencies and high values of conductivity. The difference is the dominance of eddy currents and the fact that the field from currents counteracts its source of induction due to Lenz's law.  

We also find a different scaling between both geometric cases. From Fig.\,\ref{fig:displacements}, we see that $|\,\delta_{\parallel}\,|\,\approx\,2\,\cdot\,|\,\delta_{\perp}\,|$ for the different combinations of permeability and low frequencies ($f<10^4\,\mathrm{Hz}$). This is a direct result from the observation we made in Section \uproman{2}, which is the stronger secondary field in the parallel case due to the excitation of an oscillating dipole field. For static, but electric fields, proposed dipole models of two spheres show a similar relation between the parallel and transverse displacement \cite{qizheng2002displaced}. The predicted displacements in the limit $f\,\rightarrow\,0$ appear at a maximum of $2\%$ - $5\%$ of the sphere diameter. This is another observation which agrees with previous models, that are formulated in the magnetostatic limit \cite{du2014micro}, which is intrinsically contained by our approach.
\subsection{A simplified model without displacements}
In order to evaluate the effect of the previously derived displacement we consider a simplified version of (\ref{eqn:parallel_dipole_iteration_formula}) and (\ref{eqn:perp_dipole_iteration_formula}) by setting $\delta^{(n)}_{\parallel}\!=\!\delta^{(n)}_{\perp}\!=\!0~\forall\,n\!\in\!\mathbb{N}$. This simplified version, discussed in \cite{mehdizadeh2010interaction}, will be referred to as the interacting dipole model ($ID$) with dipole field origins in the sphere centers.
The $ID$ dipoles can be written as closed expressions after taking $n\!\rightarrow\! \infty$ in (\ref{eqn:parallel_dipole_iteration_formula}) and (\ref{eqn:perp_dipole_iteration_formula}) and using the convergence of the geometric series as
\begin{align}
m_{\parallel, ID}=-\frac{2\pi R^3}{\mu_0}\frac{\alpha_1 B_0}{1-\alpha_1\left(\frac{R}{2R+a}\right)^3},
\label{eqn:ID_dipole_formula_parallel}
\end{align}
\begin{align}
m_{\perp, ID}=-\frac{2\pi R^3}{\mu_0}\frac{\alpha_1 B_0}{1+\frac{\alpha_1}{2}\left(\frac{R}{2R+a}\right)^3}.
\label{eqn:ID_dipole_formula_perp}
\end{align}
As the displacements (\ref{eqn:delta_parallel_final}) and (\ref{eqn:delta_perp_final}) additionally depend on the second-order response factor $\alpha_2$, the simplified $ID$ model\,(\ref{eqn:ID_dipole_formula_parallel}) and\,(\ref{eqn:ID_dipole_formula_perp}) is expected to be less accurate because it completely omits the second order.

The set of frequencies, at which the shift vanishes,
\begin{align}
\left\{f\in \mathbb{R}: \lim_{n\rightarrow \infty} \delta_{\parallel, \perp}^{(n)}=0 \right\},
\end{align}
tells where our proposed $IDD$ model will not provide significantly different results compared to the computationally simpler $ID$ model.
\section{Computational results}
\subsection{Comparison to FEM data}
The model is evaluated and tested by comparison to FEM simulation data\footnote{The commercial magnetic simulation software \textit{CST Studio Suite\textsuperscript{\textregistered}, version 2020}, was used.}. For a high reliability all software properties were calibrated to the case of a single sphere in an oscillating homogeneous magnetic field, such that the analytically predicted dipole moment (\ref{eqn:secondary_dipole_B0}) could be reproduced to a maximum relative error of $0.05\%$ in amplitude and phase. This was done for all considered parameter combinations.

\begin{figure}[t]
\centering
\includegraphics[width=0.45\textwidth]{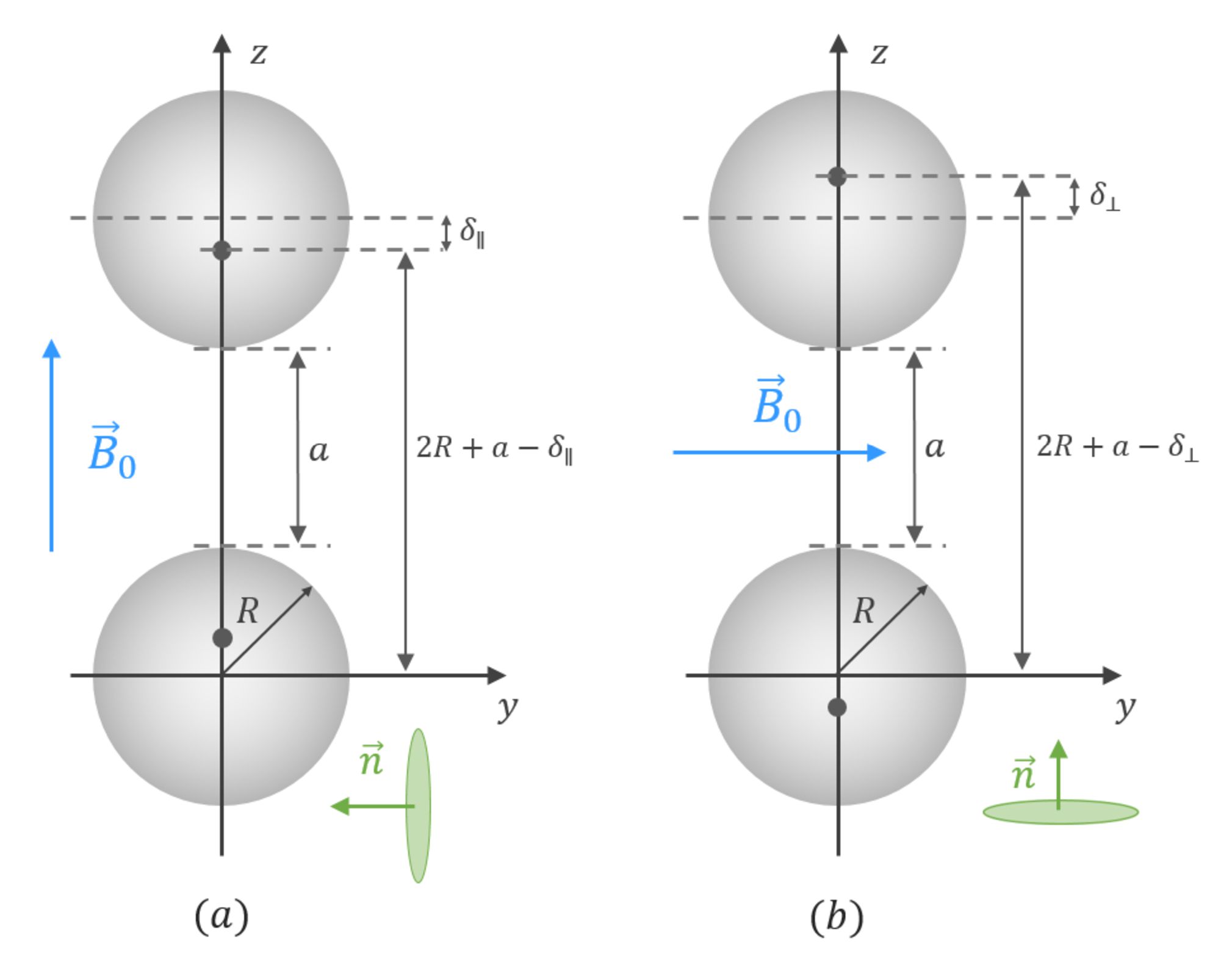}
\caption{\small{Simulation setup for parallel excitation (a) and transverse excitation (b). The sphere-to-sphere distance is $a$ and the distance between the excitation dipole field origin, marked with a dot, in sphere $2$ and the center of sphere $1$ is $2R\!+\!a\!-\!\delta_{\parallel,\perp}$. The induced secondary field is integrated over the circular loop with normal vector $\bm{n}$.}}
\label{fig:measurement_setup}
\end{figure}

The secondary magnetic field of the spheres is calculated against the distance $a$. The measurement setup is illustrated in Fig.\,\ref{fig:measurement_setup}(a) for parallel excitation and in Fig.\,\ref{fig:measurement_setup}(b) for transverse excitation, respectively.
The radius of both spheres is set to $R=10~\mathrm{mm}$. Sphere $1$ is placed at $(0,0,0)$ and sphere $2$ at $(0,0,2R\!+\!a)$. For data extraction we integrate the secondary field over a fictitious loop with radius $R/2\!=\!5~\mathrm{mm}$ at position $(0, 1.5R, -R)$ and normal vector $\bm{n}\!\perp \!\bm{B}_0$. The resulting amplitude is then a magnetic flux in the dimension of $\mathrm{nWb}$. The distance $a$ is increased starting from $a\!=0.01~\mathrm{mm}\!=\!1/100\,R$ to $a\!=\!100~\mathrm{mm} \!=\!10\,R$.
In addition to FEM data, we compare the $IDD$ model to the simplified $ID$ model and the $AD$ model, of which the latter ignores any interaction. 
\begin{figure*}[ht]
\centering
\includegraphics[width=0.75\textwidth]{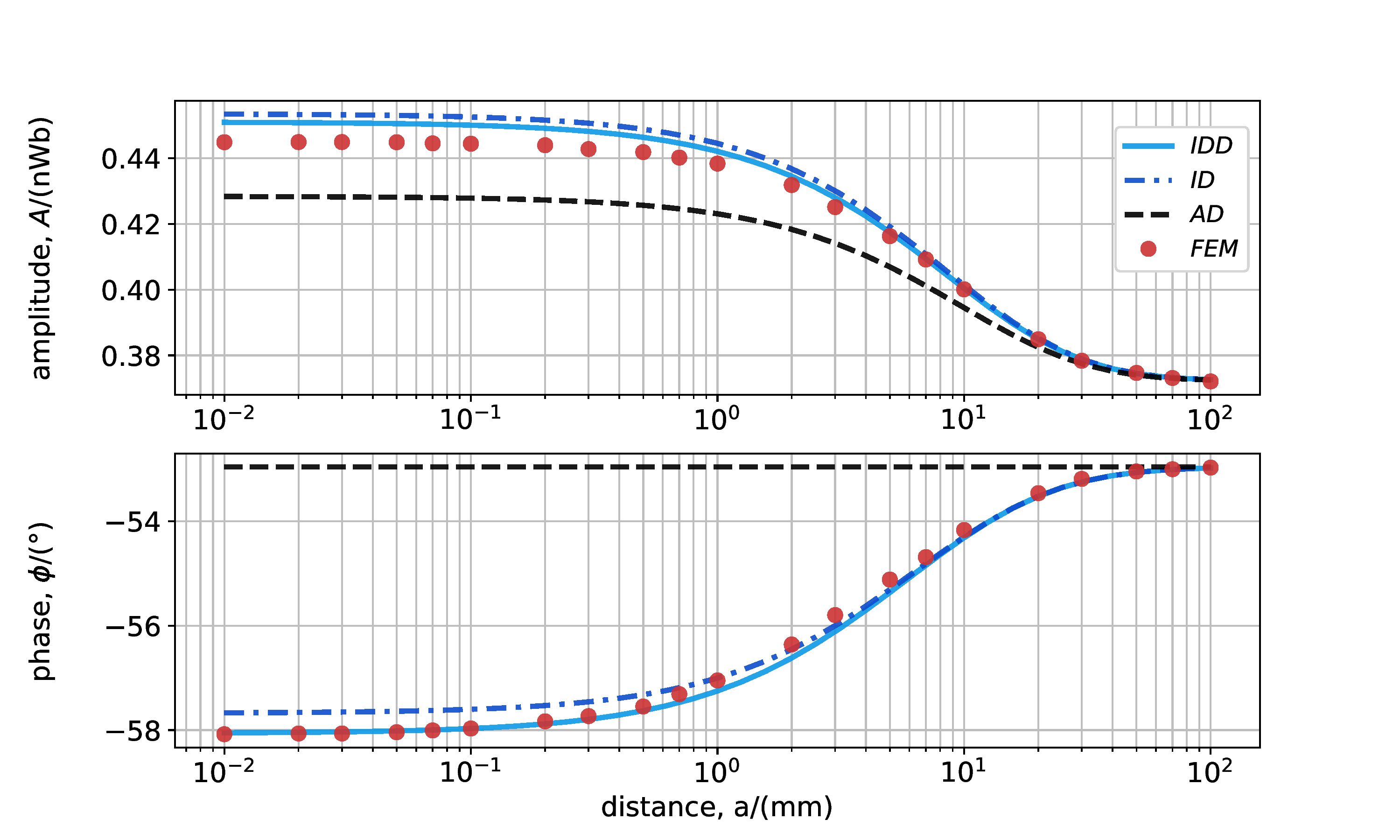}
\caption{\small{Amplitude and phase of flux through the circular loop against surface-to-surface distance $a$ for FEM simulation and model (AD, ID, IDD) data of the integrated secondary field for parallel excitation. The parameters were set to $\mu_r\,=\,73.5$, $\sigma\,=\,5\cdot\,10^{6}\,\mathrm{S/m}$, $R\,=\,10\,\mathrm{mm}$, and $f\,=\,20\,\mathrm{kHz}$.}}
\label{fig:results_two_spheres_parallel}
\centering
\includegraphics[width=0.75\textwidth]{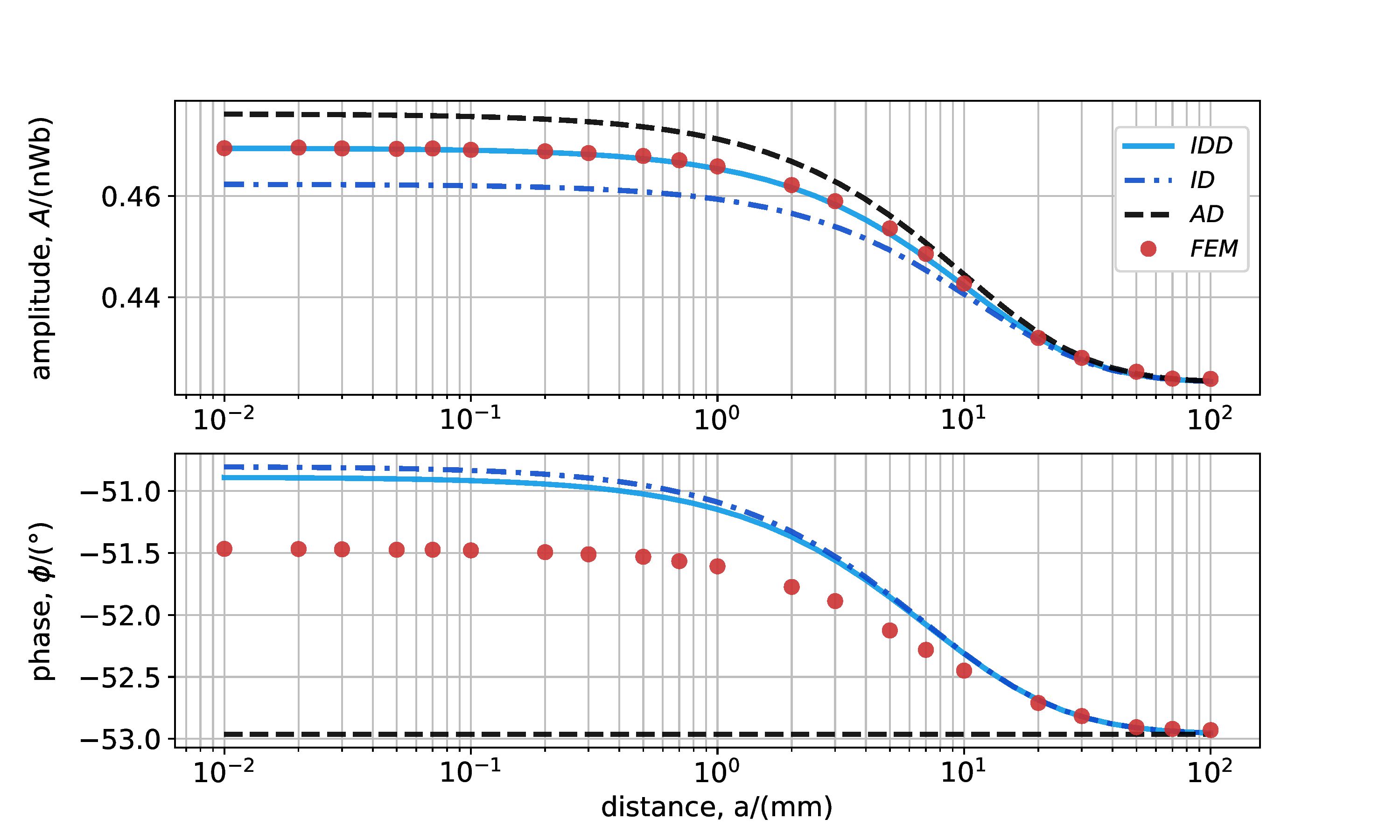}
\caption{\small{Amplitude and phase of flux through the circular loop against surface-to-surface distance $a$ for FEM simulation and model (AD, ID, IDD) data of the integrated secondary field for transverse excitation. The parameters were set to $\mu_r\,=\,73.5$, $\sigma\,=\,5\cdot\,10^{6}\,\mathrm{S/m}$, $R\,=\,10\,\mathrm{mm}$, and $f\,=\,20\,\mathrm{kHz}$.}}
\label{fig:results_two_spheres_perp}
\end{figure*}
Motivated by preceding experimental work, we set the material parameters to $\mu_r\!=\!73.5$ and $\sigma\!=\!5\cdot \!10^6~\mathrm{S/m}$. The frequency is chosen as $f\!=\!20~\mathrm{kHz}$, which is a commonly used mode of operation \cite{o2017fast}, \cite{dholu2017eddy}. 

The results of the parallel excitation can be seen in Fig.\,\ref{fig:results_two_spheres_parallel}. 
For large distances $a$, all models converge to the same values, matching the simulation data both in amplitude and phase. This reflects a decreased impact of mutual interaction for larger distances and comes from the fact that the secondary field of one sphere over the volume of the other sphere becomes weaker and more homogeneous. The $AD$ model shows, as expected, no dependence of the phase values on the distance, whereas the amplitude data of this simple model at least qualitatively follows the FEM data.
The range of phases which is covered in this scenario is approximately $5^{\circ}$.

The $ID$ model already yields an increase in accuracy in amplitude, coinciding with the results in \cite{mehdizadeh2010interaction}, as well as in phase. But especially at small distances a difference in phase of $0.4^{\circ}$ remains. This remaining gap is closed by our $IDD$ model. For small distances, a relative error of $1.3\%$ by the $IDD$ model yields slightly better accuracy in amplitude data, compared to the $1.9\%$ which are provided by $ID$.

The results of the transverse excitation, shown in Fig.\,\ref{fig:results_two_spheres_perp}, are slightly different from the previous case: first, the range of covered phase values is only $1.5^{\circ}$ and thus smaller than for parallel excitation. With increasing distance, the phase values monotonically decrease with distance $a$ in contrary to Fig.\,\ref{fig:results_two_spheres_parallel}, where the phase increases.

While the $ID$ model predicts more accurate values than when ignoring any interaction, our new model leads to improvements especially in the amplitude data. These are even more accurate than for parallel excitation. In phase data though, the $IDD$ model is only able to yield slight improvements. At small distances an error of $0.58^{\circ}$ remains. This means that our proposed $IDD$ model in the transverse case is even a little worse than $ID$ in the parallel case.

\begin{figure*}[ht]
\centering
\includegraphics[width=0.75\textwidth]{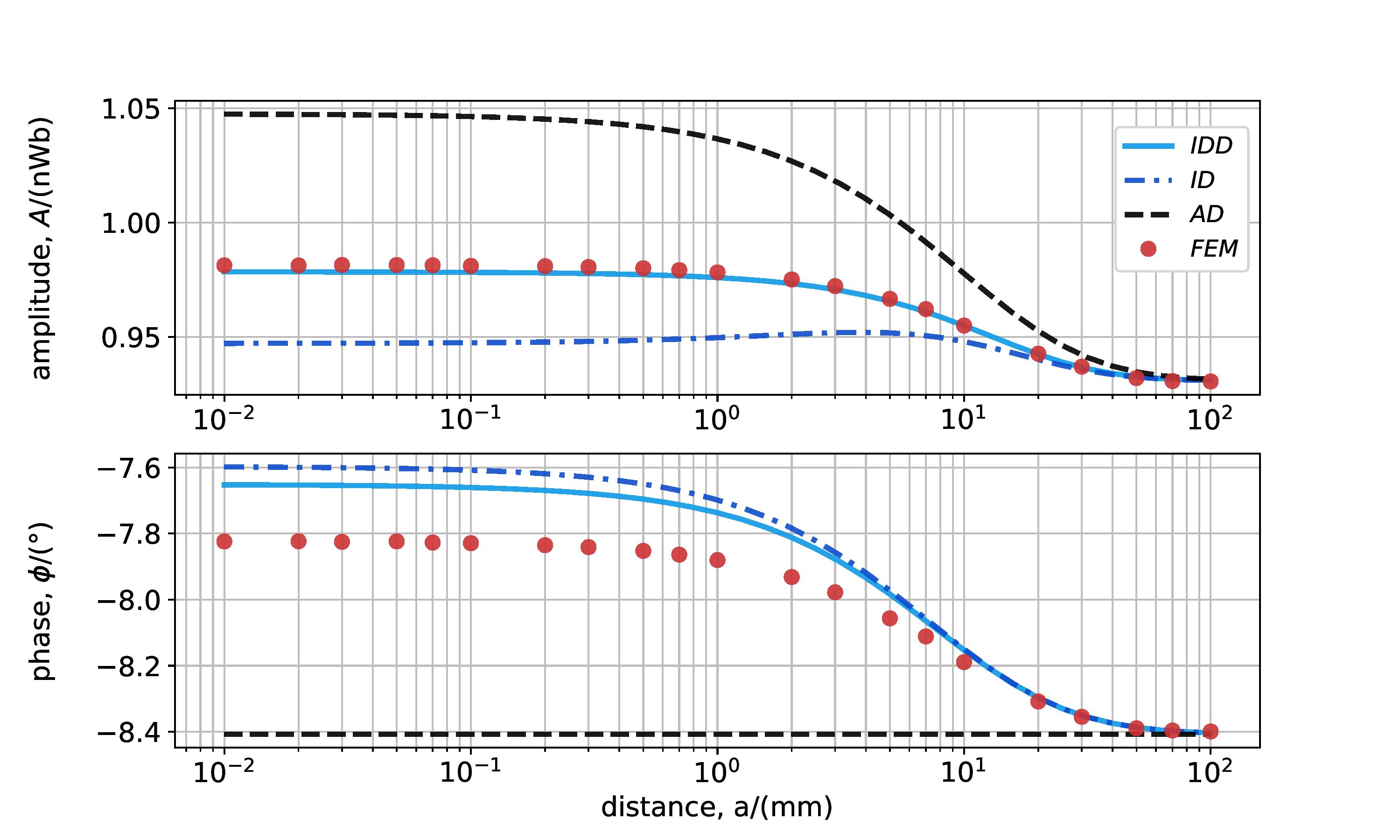}
\caption{\small{Amplitude and phase of flux through the circular loop against surface-to-surface distance $a$ for FEM simulation and model (AD, ID, IDD) data of the integrated secondary field for transverse excitation. The parameters were set to $\mu_r\,=\,73.5$, $\sigma\,=\,10^{5}\,\mathrm{S/m}$, $R\,=\,10\,\mathrm{mm}$, and $f\,=\,20\,\mathrm{kHz}$.}}
\label{fig:results_two_spheres_perp_less_sigma}
\end{figure*}

To demonstrate that the accuracy in phase data for transverse excitation depends on the material parameters, we reduce the conductivity in two steps from $\sigma\!=\!5\cdot 10^6~\mathrm{S/m}$ to a weaker  conductivity $\sigma\!=\!10^5~\mathrm{S/m}$ and to almost non-conductivity $\sigma\!=\!10^2~\mathrm{S/m}$. For both new materials the transverse excitation is again calculated. As the resulting curves look very similar and only have different scales, we solely show the results of $\sigma\!=\!10^5~\mathrm{S/m}$ in Fig.\,\ref{fig:results_two_spheres_perp_less_sigma}.
Comparing $IDD$ between the high-conductivity case, shown in Fig.\,\ref{fig:results_two_spheres_perp}, and the low-conductivity case in Fig.\,\ref{fig:results_two_spheres_perp_less_sigma}, we see that the accuracy in phase data for $a\,\rightarrow\,0$ increases from $0.58^{\circ}$ to $0.17^{\circ}$. Although the relative accuracy in phase almost stays the same, absolute phase values have more significance, as ground truth FEM data may not always be available. Also the accuracy in phase values predicted by the $ID$ model increases compared to the previous configuration. The same can be said for an almost vanishing conductivity, which is not shown here, whereas in this case the range of covered phases contracts around $0$. For both of the additionally tested conductivity values, the accuracy in amplitude data remains the same.
\subsection{Convergence of the model}
An investigation of the minimum number of iteration steps that is necessary to reach convergence is performed.
This minimum number is defined as
\begin{align}
N_i \in \mathbb{N}: |i^{(n)}-i^{(n-1)}|<\epsilon_i~\forall~n\geq N, 
\label{eqn:N_min_epsilon}
\end{align}
where $i$ is either the amplitude, the phase, or the displacement. We find that $N_i\,=\,\mathcal{O}\left(-\,\mathrm{log}~\epsilon_i\,\right)$ for a wide range of parameters $\mu_r,\,\sigma,\,\mathrm{and}\,f$. As an example of two spheres with radius $R\!=\!10~\mathrm{mm}$ at distance $a\!=\!0.1~\mathrm{mm}$, we find  $N_i\leq 25$ for $\epsilon_i=10^{-10}$ in all three quantities in both parallel and transverse excitation.
\section{Discussion and further work}  
In this paper we use the general assumption that each of the spheres exhibits an effective dipole-like secondary field from the homogeneous excitation. The more similar the overall secondary magnetic field of the spheres in reality is to a dipole field, the less error is introduced by the negligence of the $l\!>\!1$ multipole terms. Whereas this error in the data is not compensated by the ID and even less by the AD model, as expected from earlier research \cite{mehdizadeh2010interaction, wang2003frequency}, we demonstrated that our IDD model yields better accuracy. This has been done exemplarily for an oscillating excitation field, similar to the increase in accuracy that can be achieved when induced dipole field origins are shifted in the magnetostatic case \cite{du2014micro}. 

In the case of larger frequencies and highly conductive materials, currents appear within the spheres and  have circular shape only for parallel excitation but not in its transverse counterpart \cite{satpathy1971induced}. On the one hand, the resulting secondary field distribution from a transverse excitation dipole moment turns out to indeed resemble a dipole field outside the sphere, as we showed in Fig.\,\ref{fig:field_distribution_GW}(b), even for induced currents ($f\,\sigma\!\gg\!1$). On the other hand, the attendance of a second sphere, which is not present in Fig.\,\ref{fig:field_distribution_GW}(b), but in our model, may lead to additional current distortion effects that do not appear for low frequencies and non-conductive materials. These effects may make the model accuracy sensitive to the parameter product $f\,\sigma$, which controls the dominance of induced currents. From symmetry, these distortions do not change the circular shape for parallel excitation and therefore keep the assumption of dipole-like secondary fields. But in the transverse case, this statement cannot be made without further analysis and current distortion can turn the actual field less similar to a dipole field and may explain the observed phase inaccuracy from Fig.\,\ref{fig:results_two_spheres_perp}. According to this argumentation, we need to take such effects into account to further improve our model.

The investigated non-conductive scenario mathematically coincides with the case of a static excitation field as no currents will appear. In this case, the phase loses all of its information and simply approaches $0$ for all sphere distances.

From the different ranges of covered phase values, we also see that overall effects of mutual interaction for transverse excitation are not as strong as in the parallel case. This again fits the two observations: first in Section\,\uproman{2}, that the secondary field induced from an oscillating dipole field is weaker in the transverse scenario. Second, in Section \uproman{3}, where the absolute value of parallel displacement of the dipoles was larger than the transverse displacement.

An important property of our approach is that other models, which restrict themselves to the static magnetic case, like \cite{du2014micro}, are limit scenarios and intrinsically covered by our proposition.

Analog to the response factors for full spheres used in this work, there also exist similar factors for coated spheres and hollow spheres \cite{mrozynski1998analytical}. The application of our model to these cases remains future work. Also batches of more than two spheres in a homogeneous magnetic field can be modeled by our approach via iterating over pairs of spheres. In that case, in each step the fields acting on one sphere should be decomposed into parallel and transverse components for applying the model.

Furthermore, our work can serve as a basis to model more complex ensembles of metal objects. Making use of available research results for the secondary response of rotationally symmetric objects to homogeneous excitations, like small disks or spheroids \cite{norton2005eddy, mahmoud1999magnetic}, our approach may be applicable to these cases as well. This can enable the modeling of packings of more commonly used objects. Also current paths between objects, as they would appear in clusters of touching objects, are an additional phenomenon to consider.
\section{Conclusion}
In this article, we derived the IDD model for the magnetic response of two equal spheres when simultaneously excited by an oscillating homogeneous magnetic field.
The excitation was geometrically decomposed into two orthogonal cases, a parallel and transverse excitation relative to the axis through the sphere centers. The secondary magnetic field of each sphere was modeled as a single effective dipole field, which has two contributions: first, a pure dipole field caused by the primary homogeneous excitation and second, a multipole field caused by the retroactive excitation of the effective dipole field of the other sphere.

Staying with the formulation on the dipole level, we introduced an error by neglecting higher order terms in the expressions of the derived secondary magnetic dipole moments. 
To compensate for the error made with this simplification, we proposed a displacement of the origins of the secondary magnetic dipole fields out of the sphere centers. Both provided iterative expressions for the dipole moment and the displacement enabled us to accurately and quickly calculate the amplitude and phase of the secondary magnetic field of two spheres. An advantage of this new model is the applicability on static as well as oscillating magnetic fields and it can be seen as a first step towards the magnetic response prediction of geometrically manifold packings of objects, also others than spheres.

The minimum iteration number to reach convergence up to an accuracy $\epsilon$ turned out to behave as $\mathcal{O}(-\mathrm{log}\,\epsilon)$ or faster for a wide range of typical system parameters. It is therefore a preferable model to predict the magnetic two-sphere response when FEM simulations are computationally too expensive and at the same time a good accuracy with a simple model is ought to be achieved.
\appendices

\ifCLASSOPTIONcaptionsoff
  \newpage
\fi

\end{document}